\documentclass[prb,aps,nofootinbib,amssymb,twocolumn,superscriptaddress,10pt]{revtex4-2}

\usepackage[utf8]{inputenc}
\usepackage{hyperref}
\usepackage{amsmath}
\usepackage{xcolor}

\usepackage{theorem}

\newtheorem{proposition}{Proposition}

\newtheorem{theorem}[proposition]{Theorem}

\newcommand{\ket}[1]{\left|#1\right\rangle}

\def\matrix22#1#2#3#4{\left(\begin{array}{cc}#1&#2\\#3&#4\end{array}\right)}

\begin{document}

\title{Integer characteristic polynomial factorization and Hilbert space fragmentation}

\author{Nicolas Regnault}
\affiliation{Center for Computational Quantum Physics, Flatiron Institute, 162 5th Avenue, New York, NY 10010, USA}
\affiliation{Laboratoire de Physique de l'Ecole normale sup\'{e}rieure, ENS, Universit\'{e} PSL, CNRS, Sorbonne Universit\'{e}, Universit\'{e} Paris-Diderot, Sorbonne Paris Cit\'{e}, 75005 Paris, France}
\affiliation{Department of Physics, Princeton University, Princeton, New Jersey 08544, USA}

\author{Shuo Liu}
\affiliation{Department of Physics, Princeton University, Princeton, New Jersey 08544, USA}

\author{B. Andrei Bernevig}
\affiliation{Department of Physics, Princeton University, Princeton, New Jersey 08544, USA}
\affiliation{Donostia International Physics Center, P. Manuel de Lardizabal 4, 20018 Donostia-San Sebastian, Spain}
\affiliation{IKERBASQUE, Basque Foundation for Science, Bilbao, Spain}

\date{\today}

\begin{abstract}
Models with Hilbert space fragmentation are characterized by (exponentially) many dynamically disconnected subspaces, not associated with conventional symmetries but captured by nontrivial Krylov subspaces. These subspaces usually exhibit a whole range of thermalization properties, from chaotic to integrable, to quantum many-body scars. However, so far, they have not been properly defined, nor can they be easily found, given a Hamiltonian. In this work, we consider Hamiltonians that have integer representations, a common feature of many (most) celebrated models in condensed matter. We show the equivalence of the integer characteristic polynomial factorization and the existence of Krylov subspaces generated from integer vectors. Considering the pair-hopping model, we illustrate how the factorization property can be used as a method to unveil Hilbert space fragmentation. We discuss the generalization over other rings of integers, for example those based on the cyclotomic field which are relevant when working in a given ($\ne 0, \pi$) momentum sector.
\end{abstract}

\maketitle

\emph{Introduction.} Thermalization, or its absence, in out-of-equilibrium quantum many-body systems has been rendered ``laboratory"-possible with the recent progress in the experimental development of quantum simulators. The paradigm of the Eigenstate Thermalization Hypothesis (ETH)~\cite{deutsch1991quantum,srednicki1994chaos} has been thought to govern the dynamics of these quantum systems by characterizing all the excited states in the middle of the energy spectrum, in absence of integrability. However, violations of ETH beyond integrable systems have been shown to exist. They include many-body localization (MBL) where the alliance of strong disorder and interaction leads to emergent integrability (see Refs.~\cite{abanin2018review,nandkishore2015many} for comprehensive reviews). More recently, the concept of Quantum Many-Body Scars (QMBS) has been introduced from both experimental~\cite{turner2017quantum, turner2018quantum} and theoretical~\cite{moudgalya2018a,moudgalya2018b} motivations: several frameworks~\cite{mori2017eth, schecter2019weak, chattopadhyay2019quantum, mark2020unified, moudgalya2020eta, pakrouski2020many, ren2020quasisymmetry, odea2020from,moudgalya2022} now show the ubiquity of non-thermal excited eigenstates in an otherwise background of thermal eigenstates (see Refs.~\cite{SerbynReview,papic2021weak,Moudgalya2022review,Chandran2022} for reviews).

QMBS can be seen as a special case of the more generic phenomenon of ``Hilbert space fragmentation"~\cite{sala2019ergodicity,khemani2019local,moudgalya2021hilbert}. There, the Hilbert space can be decomposed into (sometimes, depending on whether the fragmentation is strong) exponentially many dynamically disconnected subspaces that are not captured by conventional symmetries. Such feature in a model potentially unveils a rich structure with subspaces that might have different thermalization properties \cite{moudgalya2019thermalization}. Fragmentation was explicitly pointed out in the context of dipole-moment or center-of-mass conserving systems~\cite{pai2018localization, sala2019ergodicity, khemani2019local, moudgalya2019thermalization}, although similar phenomena have been discussed in several works~\cite{ritort2003glassy, bergholtz2005half, bergholtz2006one, olmos2010thermalization, sikora2011extended, nakamura2012exactly, gopalakrishnan2018facilitated, lan2018quantum, moudgalya2020quantum,Yang2020,Chen2021,Pozsgay2021,lehmann2022,frey2022,nicolau2022,pozderac2022,Bhattacharjee2022}.

Hilbert space fragmentation can be framed in a more mathematical manner using Krylov subspaces. A (maximal) Krylov subspace is spanned from a state by repeated action of the Hamiltonian. In a fragmented system, the Hilbert space decomposes as a direct sum of an exponential number of Krylov subspaces. It immediately raises the question - and the apparent paradox in answering it - of which states should be used to define these subspaces. A random many-body state would typically lead to a Krylov subspace identical to the whole Hilbert space. As an extreme case, considering the eigenstates would trivially give as many Krylov subspaces as the total Hilbert space dimension. In most of the Hilbert space fragmentation examples, these states are many-body product states, up to potential symmetries, that also turn out to be those that are relevant as initial states for quantum simulators. Finding which states lead to nontrivial fragmentations is usually challenging and model dependent. It is clear that many aspects about fragmentation remain vague or
unanswered. This led Ref.~\cite{moudgalya2021hilbert} to define Hilbert space fragmentation only for \emph{classes} of Hamiltonians rather than for specific ones, for which there is still no clear definition, despite concrete examples. 

In this article, we intend to provide a different - descriptive and prescriptive - perspective on Hilbert space fragmentation based on characteristic polynomials rather than finding states for Krylov subspaces. For that purpose, we consider a specific class of Hamiltonians admitting integer matrix representations, and as such integer characteristic polynomials. While sounding restrictive at first, this hypothesis is actually satisfied by many (most) celebrated models in physics such as the Heisenberg spin model, the Affleck-Kennedy-Lieb-Tasaki (AKLT) model~\cite{aklt1987rigorous}, Hubbard model at $U/t$ rational, or the pair-hopping model~\cite{seidel2005incompressible, moudgalya2020quantum, moudgalya2019thermalization}, an example of interacting dipole conserving model exhibiting Hilbert space fragmentation. We show that the integer factorization of the characteristic polynomial of the Hamiltonian (beyond that given by symmetries) is \emph{equivalent to the existence of Krylov subspaces spread by integer vectors}. Illustrating this property through the pair-hopping model, we show that the integer characteristic polynomial factorization provides a unique approach to test if a model is subject to Hilbert space fragmentation. We also discuss generalizations to other rings of integers when dealing with, e.g., symmetries such as translations. Our paper gives both a definition and a clean mathematical structure as well as a very practical method to search for Hilbert space fragmentation in Hamiltonians.

\emph{Krylov subspace and factorization.} Let $H$ be a (Hermitian) Hamiltonian and $\cal H$ the Hilbert space of dimension $\cal N$. The Krylov subspace $\cal{K}$ generated by a state $\ket{\Psi_0}$ and of dimension $n$ is the subspace spanned by the linearly independent set of vectors $\left\{\ket{\Psi_0}, H \ket{\Psi_0},..., H^{n-1} \ket{\Psi_0}\right\}$. Note that these vectors are usually neither normalized nor orthogonal, even if $\ket{\Psi_0}$ was normalized. In this article, we will consider only \emph{maximal} Krylov subspace, i.e., we assume that $H^{n}\ket{\Psi_0}$ belongs to $\cal{K}$ and $n < \cal{N}$. As a consequence, the Hamiltonian has a block diagonal structure when decomposing the Hilbert space into ${\cal H}={\cal K} \oplus \bar{\cal K}$, where $\bar{\cal K}$ is the complement of ${\cal K}$ in $\cal H$.

We now assume that $H$ has an integer representation in a given orthonormal basis ${\cal B}=\left\{\ket{\phi_j}\right\}$ of $\cal{H}$. Consider a state $\ket{\Psi_0}$ that has an integer decomposition onto $\cal B$ and generates a Krylov subspace $\cal{K}$ of dimension $n$. We will now prove that the existence of such a Krylov subspace implies that the characteristic polynomial of $H$ admits a factorization over the integer polynomial. First, note that each vector $H^i\ket{\Psi_0}$ is also an integer vector when decomposed on $\cal{B}$. By definition of $\cal{K}$, $H^n\ket{\Psi_0}$ has a unique decomposition
\begin{eqnarray}
  H^n\ket{\Psi_0} &=& \sum_{i=0}^{n-1} v_i H^i \ket{\Psi_0}\label{eq:DefVCoefficents}
\end{eqnarray}
where $v_i$ are coefficients. Being a linear relation between integer vectors, these $v_i$'s have to be either integers or rational numbers. We define the transformation matrix $U$ as

\begin{eqnarray}
U&=&\left[\ket{\Psi_0}, H \ket{\Psi_0},..., H^{n-1} \ket{\Psi_0}\right]\label{eq:DefUMatrix}
\end{eqnarray}

\noindent built from the linearly independent states spanning $\cal{K}$. Using Eq.~\ref{eq:DefUMatrix}, we have

\begin{eqnarray}
HU&=&U\left(\begin{array}{ccccc|c} 0 & 0 & 0 & ... & 0 & v_0 \\ 1 & 0 &  0 & ... & 0 & v_1 \\ 0 & 1 &  0 & ... & 0 & v_2 \\ \vdots & \vdots & \vdots & \vdots & \vdots \\ 0 & 0 &  0 & ... & 1 & v_{n-1} \end{array}\right).\label{eq:HOnU}
\end{eqnarray}

Since the $H^i \ket{\Psi_0}$, $0\le i<n$ are linearly independent states, there exists a ${\cal{N}} \times n$ matrix $V$ such that $V^{t}U=I_{n}$, where $I_{n}$ is the $n \times n$ identity matrix. Thus the projection of the Hamiltonian on $\cal{K}$ reads

\begin{eqnarray}
V^{t}HU &=& \left(\begin{array}{ccccc|c} 0 & 0 & 0 & ... & 0 & v_0 \\ 1 & 0 &  0 & ... & 0 & v_1 \\ 0 & 1 &  0 & ... & 0 & v_2 \\ \vdots & \vdots & \vdots & \vdots & \vdots \\ 0 & 0 &  0 & ... & 1 & v_{n-1} \end{array}\right).\label{eq:HInKrylov}
\end{eqnarray}

Its characteristic polynomial $P_{\cal K}(\lambda)$ can be directly evaluated by developing the determinant with respect to the last column containing the $v_i$ coefficients, namely

\begin{eqnarray}
P_{\cal K}(\lambda)&={\rm det}\left(\lambda I_{n} - V^{t}HU\right) &= \lambda^n - \sum_{i=0}^{n-1} v_i \lambda^i. \label{eq:CharPolynomialKrylov}
\end{eqnarray}
Indeed, the minor of $v_i$ is the determinant of a lower triangular matrix with a diagonal of $(n-1-i)$ $-1$'s and $i$ $\lambda$'s. Per Eq.~\ref{eq:CharPolynomialKrylov}, we deduce that the characteristic polynomial $P_{\cal K}(\lambda)$ of $H$ restricted to the Krylov subspace ${\cal K}$ is a \emph{rational} polynomial. 

The characteristic polynomial $P(\lambda)$ of $H$ trivially is a (primitive) integer polynomial, since $H$ has an integer representation. Let denote $P_{\bar{\cal K}}(\lambda)$ the characteristic polynomial of the projection of $H$ on $\bar{\cal K}$. Thanks to the block structure of $H$ in ${\cal K} \oplus \bar{\cal K}$, $P(\lambda)$ factorizes as 

\begin{eqnarray}
P(\lambda)&=&P_{\cal K}(\lambda) P_{\bar{\cal K}}(\lambda)\label{eq:PolynomialFactorization}
\end{eqnarray}
meaning that $P_{\bar{\cal K}}(\lambda)$ is also a rational polynomial and $P(\lambda)$ admits a factorization over the rational polynomials. But Gauss's lemma (over unique factorization domains such as the integers) implies that if a primitive integer polynomial has a factorization over rational polynomials, then these rational polynomials are actually integer polynomials. We conclude with the following Theorem~\ref{theorem:thm1}:

\begin{theorem}\label{theorem:thm1}
The existence of a nontrivial Krylov subspace from an integer vector for an integer Hamiltonian leads to an integer factorization of its characteristic polynomial. 
\end{theorem}

\emph{Applications.}  A trivial application of Theorem.~\ref{theorem:thm1}
corresponds to the situation where $\ket{\Psi_0}$ is an eigenstate related to an integer eigenvalue $\lambda_0$. It can easily be shown by Gaussian elimination that any integer eigenvalue of an integer matrix has a rational eigenstate and thus by multiplying the product of denominators, an integer eigenstate. As such, the characteristic polynomial can be factorized by the degree one polynomial $\lambda-\lambda_0$.

A more interesting application concerns symmetries. From now on, we consider that the basis $\cal B$ is a physically motivated basis, typically a product state basis such as quantum spins on a lattice, Fock states for sites and orbitals, etc.  This basis might be resolved in a given quantum number sector from, e.g., a $U(1)$ symmetry like the total spin projection or the charge conservation, as long as such a resolution preserves the product state nature of $\cal B$. This symmetry trivially implies the factorization of the characteristic polynomial.  Let us assume that the system has another symmetry not directly encoded in the basis $\cal B$. For sake of concreteness, we focus on one dimensional discrete translation invariance for a system with $L$ sites or unit cells. We denote by $T$ the translation operator for one site and we suppose that $\ket{\psi_0}$ is an eigenstate of $T$. If its momentum $k$ is $0$ or $\pi$, coefficients can be chosen to preserve the property of being integers: configurations related by $n$ translations $T^n$ should be identical up to a factor $e^{ikn}=1$ or $-1$. Apart from fine tuning, any random choice of $\ket{\psi_0}$ spans a Krylov subspace that matches the Hilbert space subspace with the same quantum numbers as $\ket{\psi_0}$. Using Theorem.~\ref{theorem:thm1},
we deduce that the characteristic polynomial factorizes into integer polynomials for the $k=0$, $k=\pi$ and another polynomial for the all other sectors. Note that this latest might also exhibit further factorizations: for example when combining $k=\pm\frac{\pi}{2}$ sectors, if the system also has inversion symmetry, the polynomial will further factorize over the integers. We could also derive the same results through basis transformation by combining configurations related by translations in $\cal B$ into eigenstates of $T$ with momentum $k=0$ or $\pi$. Once applied to the Hamiltonian, such a transformation leads to another integer Hamiltonian albeit non-symmetric in general (due to the different orbit size of configurations under translations).

Other symmetries could also lead to a similar integer polynomial factorization. For example, spin systems with $SU(2)$ symmetry made of either spins $1/2$ or spin $1$ would also exhibit an integer factorization of their characteristic polynomial per total spin sector. Consider the projector onto the total spin $s$ defined as
\begin{eqnarray}
{\cal P}_s&=& \prod_{j \neq s} \frac{j(j+1) - S^2}{j(j+1)-s(s+1)}\label{eq:spinprojector}
\end{eqnarray}
where $S^2$ is the total spin operator and $j$ runs over all the possible total spin except $s$. In the spin product state basis,  $S^2$, and thus ${\cal P}_s$, admits an integer representation for either spins $1/2$ or spin $1$. Assuming the total Hamiltonian $H$ has an integer representation in this basis, its projection ${\cal P}_s H {\cal P}_s$ onto the total spin $s$ also possesses an integer representation and, as a consequence, an integer characteristic polynomial. Using the $SU(2)$ symmetry, we deduce that the characteristic polynomial of an $SU(2)$ symmetric $H$ factorizes over integers with factors associated to every total spin sector. As we will show below, factorization also means that we can find an integer vector with a fixed total spin generating a Krylov subspace associated to the corresponding integer polynomial factor. As such, it is related to nontrivial structure beyond symmetries, such as Hilbert space fragmentation.

\emph{Nontrivial factorizations.} Once integer factorizations of the characteristic polynomial due to symmetries have been taken out (this later will be extended over the field of cyclotomic integers), any further integer polynomial factorization is \emph{a nontrivial property} of the model. As shown previously, any Krylov subspace generated from an integer vector and not identical to a whole Hilbert subspace with fixed quantum numbers, would induce such nontrivial factorization. Conversely, 
\begin{theorem}\label{theorem:thm2}
A nontrivial factorization can be associated to a Krylov subspace.
\end{theorem}
Let us denote by $P_{\cal \kappa}(\lambda)$ the polynomial factor of degree $n$, using the same monomial expansion in powers of $\lambda$ as Eq.~\ref{eq:CharPolynomialKrylov} (i.e. we know the $v_j$'s). Since $H$ is Hermitian, there exist orthogonal eigenstates $\{\ket{\varphi_i}\}$ with $i=0,...,n-1$, $n\leq {\cal N}$ associated to the roots $\{\lambda_i\}$ (i.e. eigenvalues) of $P_{\cal \kappa}(\lambda)$. We define the basis $\{\ket{e_j}\}$ with $j=0,...,n-1$ as
\begin{eqnarray}
\ket{e_j}&=& \sum_{i=0}^{n-1} \lambda_i^j  \ket{\varphi_i}.\label{eq:ejbasis}
\end{eqnarray}
Using the eigenstate definition and the identity $\lambda^n = \sum_{i=0}^{n-1} v_i \lambda^i$ (with $v_i$ integers) from Eq.~\ref{eq:CharPolynomialKrylov}, these states satisfy
\begin{eqnarray}
&&H\ket{e_j}=\ket{e_{j+1}} \;\; {\rm for}\;\; j < n-1 \label{eq:ejproperty1}\\
&&H\ket{e_{n-1}}=\sum_{j=0}^{n-1} v_j \ket{e_j}.\label{eq:ejproperty2}
\end{eqnarray}
From these relations, we can define a transformation matrix $U$ to the $\{\ket{e_j}\}$ basis that also verifies Eq.~\ref{eq:HOnU}. Note that from the linear independence of the eigenstates, we can also define a matrix $V$ such that $V^t U = I_n$, allowing to write Eq.~\ref{eq:HInKrylov}. Since both $H$ and the matrix $V^t H U$ are integer matrices (the coefficients $v_i$ being integers), we can use Eq.~\ref{eq:HOnU} and Gauss elimination to show that $U$ admits a rational expression and thus up to a global factor, an integer expression. As such, the first column of $U$ provides an integer vector $\ket{\Psi_0}$ generating the Krylov subspace related to the nontrivial factorization. We stress that this proof does not claim any property about $\ket{\Psi_0}$ such as a sparse structure in the original basis ${\cal B}$. But as we will illustrate below through the pair-hopping model, we argue that such an integer polynomial factorization is in general enough to \emph{recover the whole relevant Hilbert space fracture} (namely, there is no further fragmentations that would arise from Krylov subspaces generated from inherently irrational vectors). 

\emph{Pair-hopping model.} To show the relation between fragmentation and integer polynomial factorization, we consider a simple model of dipole-moment conserving system: the one-dimensional spinless fermionic pair-hopping model~\cite{seidel2005incompressible, moudgalya2020quantum, moudgalya2019thermalization}. This model can be seen as the quantum part of the model Hamiltonian for the Laughlin $\nu = 1/3$ fractional quantum Hall state~\cite{lee2015geometric} where the electrostatic terms are discarded and in the thin torus limit~\cite{lee2015geometric, moudgalya2020quantum}. Assuming periodic boundary conditions and $L$ sites, the pair-hopping model Hamiltonian reads

\begin{eqnarray}
H_{\rm p-h}&=& -\sum_{j = 0}^{L-1}{\left(c^\dagger_j c^\dagger_{j+3} c_{j + 2} c_{j + 1} + h.c.\right)},\label{eq:pairhopping}
\end{eqnarray}
where $c_j$ is the spinless fermionic annihilation operator on site $j$. We note $n_j$ the occupation on site $j$. Irrespective of the boundary conditions, the model has two conserved quantities: the total charge $Q=\sum_j n_j$ and the dipole moment $D=\sum_j j n_j$. Moreover when $L$ is even, the system has an additional symmetry: the sublattice particle conservation $Q_e=\sum_{j\;{\rm even}} n_j$. If periodic boundary conditions are used, $D$ is only defined modulo $L$ and a purely many-body quantum number $\kappa=0,...,{\rm GCD}(Q,L)-1$, akin to the many-body momentum due to magnetic translations in the fractional quantum Hall effect~\cite{Haldane1985},  is also conserved.

The pair-hopping model exhibits Hilbert space fragmentation ranging from Krylov subspaces of order one with respect to the system size~\cite{sala2019ergodicity,khemani2019local} up to Krylov subspaces growing exponentially as $\alpha^L$ with $1 < \alpha < 2$~\cite{moudgalya2019thermalization} (whereas the Hilbert space grows like $2^L$). Interestingly, these Krylov subspaces are generated from single product states (or a sum of one product state and its partner under the magnetic translation-like symmetry if we are interested in the $\kappa=0$ sector). The exponentially-large Krylov subspaces can be integrable, related to the spin-$1/2$ XX model, or non-integrable \cite{moudgalya2019thermalization}(either ETH or many-body localized upon addition of disorder~\cite{herviou2021manybody}).

All these Krylov subspaces mentioned previously arise from a single state of the Slater basis. Each Slater configuration can be represented as a string of 0's and 1's corresponding to the occupation of each site. The Hamiltonian Eq.~\ref{eq:pairhopping} connects any pattern $1\;0\;0\;1$ within a configuration to $0\;1\;1\;0$ and vice versa. Thus any configuration not containing any of these two patterns, dubbed frozen configuration, is annihilated by the Hamiltonian, i.e., an eigenstate with zero energy~\cite{sala2019ergodicity,khemani2019local}. Such a zero energy corresponds to a one dimensional Krylov subspace and thus is associated to a degree one factor in the integer characteristic polynomial factorization. The Krylov subspace related to the spin-$1/2$ XX model is produced by a repeated pattern $1\;0\;0\;1$. In addition, any pattern of at least three consecutive occupied sites $1\;1\;1\;...$ (or zeros) is inert upon action of Eq.~\ref{eq:pairhopping} and acts as blockade, effectively cutting the system into disconnected parts. We can combine these blockades and the configurations from, e.g., frozen configurations or XX integrable sector to generate new Krylov subspaces~\cite{moudgalya2019thermalization}.

The pair-hopping Hamiltonian Eq.~\ref{eq:pairhopping} in the fermionic Slater basis is naturally represented by an integer matrix. The abundance of Krylov subspaces provides a perfect playground for the theorems that we have proved in this article. For pedagogical purposes, we consider a small system with only $L=12$ sites at half filling, i.e., $Q=6$. We focus on a single quantum number sector, with $Q_e=3, \kappa=0, D=9$, which corresponds to the largest Hilbert subspace of dimension $12$ for this system size at half filling. Note that working in the $\kappa=0$ sector does not spoil the integer representation, as explained previously for the translation symmetry. Moreover, when we will provide a product state configuration to describe Krylov subspace, it should be understood as an equal weight superposition of all its symmetry partner in order to be invariant under this symmetry. The characteristic polynomial for the Hamiltonian in this single quantum number sector reads out 
\begin{eqnarray}
P(\lambda) &=& P_1(\lambda) P_2(\lambda) P_3(\lambda)P_4(\lambda)\label{eq:pairhoppingintegerfactorization}
\end{eqnarray}
where
\begin{eqnarray}
P_1(\lambda)&=&\lambda^2\label{eq:pairhoppingintegerfactorization1}\\
P_2(\lambda)&=& \lambda(\lambda - 4) (\lambda + 2)\label{eq:pairhoppingintegerfactorization2}\\
P_3(\lambda)&=& (\lambda^2 - 1)  (\lambda^2 - 5)\label{eq:eq:pairhoppingintegerfactorization3}\\
P_4(\lambda)&=& \lambda^2 (\lambda + 2)\label{eq:pairhoppingintegerfactorization4}. 
\end{eqnarray}
We can relate these different factors to different Krylov subspaces. $P_1(\lambda)$ corresponds to the two frozen product state configurations $\ket{111110000001}$ and $\ket{111101000010}$. $P_2(\lambda)$ is associated to the XX model integrable Krylov subspace (namely $\ket{100110011001}$), while $P_3(\lambda)$ is related to $\ket{111100100100}$ which can be seen as one blockade and a XX model string. Although $P_3(\lambda)$ and other $P_n(\lambda)$ have further factorizations at this size, we show below that this is specific to the very small dimension of this example. $P_4(\lambda)$ is the remainder of the integer factorization. For this latest factor and using our theorem, we can find an associated Krylov subspace generated from an integer vector. Actually, since all the roots of $P_4(\lambda)$ are integers, they can be associated to integer eigenvectors namely $\ket{010110011010}+\ket{011001011010}$ (with eigenvalue $-2$), 

$\ket{110111001000}+\ket{111011000100}$ and 
$\ket{111001110000}-\ket{111100011000} + \ket{111011000100 }$ (both with eigenvalues $0$, see App.~\ref{app:details}). 

For the $L=12$, $Q_e=3, \kappa=0, D=9$, Eqs.~\ref{eq:pairhoppingintegerfactorization2}-\ref{eq:pairhoppingintegerfactorization4} have several additional factorizations (such as in $P_3(\lambda)$) due to the presence of integer eigenvalues. As such, we could regroup the factors in other ways than Eq.~\ref{eq:pairhoppingintegerfactorization} and find corresponding Krylov subspaces. But in general and in absence of specific structures, such as quantum scars, the number of integer eigenvalues quickly reduces when considering larger systems, in line with the rarity of integer eigenvalues for integer matrices~\cite{Martin2009}. For example, ramping up the number of sites from $L=12$ to $L=16$, the analogue of $P_3$ (with a generating configuration having a single block of four consecutive $1$'s) corresponds to a Krylov subspace of dimension $19$ and an associated characteristic polynomial 
\begin{eqnarray}
P_3(\lambda)&=&\lambda(\lambda^8 - 15\lambda^6 + 56\lambda^4 - 61\lambda^2 + 15)\label{eq:p3forl16}\\
&& \times (\lambda^{10} - 25\lambda^8 + 188\lambda^6 - 519\lambda^4 + 479\lambda^2 - 72).\nonumber
\end{eqnarray}
Compared to Eq.~\ref{eq:p3forl16}, we now get a single integer root in addition to two integer polynomial factors of degree $8$ and $10$, respectively. This factorization is actually a consequence of the particle-hole (p-h) symmetry at half-filling: the factor of degree $10$ corresponds to one parity sector of this symmetry and the remaining polynomial to the other parity. Note that we could directly work in a given parity sector of the p-h symmetry by either projecting the Hamiltonian or considering the Krylov subspaces generated by $\ket{1111001001100100}\pm\ket{0000110110011011}$, i.e., eigenstates of the p-h symmetry. Moving to $L=20$, the analogue Krylov subspace has a dimension $81$ with the same structure: a single integer (zero energy) eigenvalue and two integer polynomial factors of degree $40$. More than the rarity of integer roots, this case also indicates that, once all symmetries are taken into account, further integer factorizations within one of the Hilbert space fragments is an artefact due to small sizes. However, our method still identifies the ``true"- thermodynamic limit, size independent Krylov subspace starting from the root $\ket{11110010010010010...}$.

\emph{Generalization and conclusion.} The mathematical derivations can be trivially extended from integers to other rings of integers as long as they admit unique factorization (domains) for the Gauss' lemma to hold. Such a generalization is helpful when dealing with, e.g., translation symmetries at certain momentum. For example, an integer Hamiltonian projected on momentum $k=\pm \frac{\pi}{2}$ leads to a matrix where entries are Gaussian integers $\mathbb{Z}[i]$, i.e., $r+is$ where $r$ and $s$ are integers. We have verified the theorems by projecting the Hamiltonian of pair-hopping model on momentum $k=\pi/2$ (see App.~\ref{app:pi2}).
Other momentum sectors $k=2\pi/n$ ($n$ being an integer) are related to the ring of integers $\mathbb{Z}[e^{2i\pi/n}]$  of cyclotomic field. Unique factorization over $\mathbb{Z}[e^{2i\pi/n}]$ only holds for certain values of $n$, for example $n=1,...,22$ but not $n=23$. A full list of the unique factorization domains is related to Fermat's theorem and is known. If the unique factorization holds, characteristic polynomial factorization over $\mathbb{Z}[e^{2i\pi/n}]$  would thus be related to the existence of nontrivial Krylov subspaces within momentum sector $k=2\pi/n$.

In this article, we proposed that nontrivial integer factorization of the characteristic polynomial for models having an integer Hamiltonian representation provides a direct method to search for Hilbert space fragmentation in a model. Our conjecture is grounded in the mathematical equivalence between maximal Krylov subspaces and integer characteristic polynomial factorization. We use the pair-hopping model as a playground to discuss this method. But other models can be investigated in a similar manner. For example, the AKLT model (for system sizes up to $12$ sites) does not exhibit any integer factorization beyond the tower of states, i.e., the AKLT QMBS, and the eigenstates with integer energies discussed in Refs.~\cite{moudgalya2018a,moudgalya2018b}. As such, it illustrates that we can use our approach to find QMBS. Even integrable models, such as the spin-$1/2$ Heisenberg model, can also be revisited with our approach, through nontrivial factorizations~\cite{upcomingwork}.

\emph{Acknowledgments.}
 N.~R. was supported by the DOE Grant No. DE-SC0016239  and by the Princeton Global
Network Funds. This work is also
partly supported by a project that has received funding
from the European Research Council (ERC) under the
European Union’s Horizon 2020 Research and Innovation
Programme (Grant Agreement No. 101020833). The Flatiron Institute is a division of the Simons Foundation.

\bibliographystyle{apsrev4-2}
\bibliography{fracturefactorization.bib}

\onecolumngrid
\pagebreak

\cleardoublepage

\appendix

\section{Details of the integer factorization for pair-hopping model}\label{app:details}
In the main text, we have discussed the nontrivial integer factorization for the specific quantum number sector $Q_{e}=3, \kappa=0, D=9$ sector at half filling with $L=12$ sites. This nontrivial integer factorization is shown in Eqs.~(12)-(16), indicating the existence of Hilbert space fragmentation. In this section, we provide details of this factorization.

The Hilbert space dimension of this quantum number sector is $12$ and spanned by the following many-body basis states
\begin{eqnarray}
    \vert \psi_{1} \rangle = \vert 111110000001 \rangle,  \\
    \vert \psi_{2} \rangle = \vert 111101000010 \rangle,  \\
    \vert \psi_{3} \rangle = \vert 100110011001 \rangle,  \\
    \vert \psi_{4} \rangle = \vert 010110011010 \rangle,  \\
    \vert \psi_{5} \rangle = \vert 011001011010 \rangle,  \\
    \vert \psi_{6} \rangle = \vert 101010010101 \rangle,  \\
    \vert \psi_{7} \rangle = \vert 111100100100 \rangle,  \\
    \vert \psi_{8} \rangle = \vert 111011000100 \rangle,  \\
    \vert \psi_{9} \rangle = \vert 111100011000 \rangle,  \\
    \vert \psi_{10} \rangle = \vert 110111001000 \rangle,  \\
    \vert \psi_{11} \rangle = \vert 101100001101 \rangle,  \\
    \vert \psi_{12} \rangle = \vert 111001110000 \rangle.  
\end{eqnarray}
As mentioned in the main text, the product state configuration should be understood as an equal-weight superposition of all its symmetry partners in order to be invariant under the discrete translation symmetry. For example, the vector $\vert \psi_{3} \rangle$ actually reads
\begin{eqnarray}
    \vert \psi_{3} \rangle &= \vert 100110011001 \rangle &\equiv \vert 100110011001 \rangle - \vert 011001100110 \rangle. 
\end{eqnarray}
Here the minus sign is induced by translation operator that translates all fermions by $L/Q=2$ sites. Similarly the expended expression of $\vert \psi_{1} \rangle$ is
\begin{eqnarray}
\vert \psi_{1} \rangle &= \vert 111110000001 \rangle &\equiv \vert 111110000001 \rangle - \vert 011111100000 \rangle - \vert 000111111000 \rangle - \vert 000001111110 \rangle \nonumber\\
&&+ \vert 100000011111 \rangle + \vert 111000000111 \rangle.
\end{eqnarray}
We note that $\vert \psi_{1} \rangle, \cdots, \vert \psi_{12} \rangle$ are unnormalized. Within this quantum number sector, the Hamiltonian matrix in the normalized basis is given by
\begin{align}
   H = \begin{pmatrix}
       H_{1} & 0 \\
       0  & H_{2} 
   \end{pmatrix},
\end{align}
where 
\begin{align}
    H_{1} = \begin{pmatrix}
    0 & & & & &    \\
     & 0  & & & & \\
     & & 0 &  -\sqrt{3} & \sqrt{3} & 0  \\
     && -\sqrt{3} & 0 & -2 & 1  \\
     && \sqrt{3} & -2 & 0 & -1  \\
     && 0 & 1 & -1 & 0
    \end{pmatrix},\label{eq:H1}
\end{align}
and 
\begin{align}
    H_{2} = \begin{pmatrix}
     0 & -1 & -1 & 1 & 0 & 0 \\
     -1 & 0 & 0 & 0  & -1 & 0 \\
     -1 & 0 & 0 & 0  & 0 & 0 \\
     1& 0 & 0 & 0  & 1& 0 \\
     0 & -1 & 0 & 1 & 0 & 1 \\
     0 & 0 & 0 & 0  & 1& 0
        \end{pmatrix}.
\end{align}

Note that we have omitted zeroes in Eq.~\eqref{eq:H1} for readability. Consequently, the characteristic polynomial within this quantum number sector is 
\begin{eqnarray}
    P(\lambda) = P_{1}(\lambda) P_{2}(\lambda) P_{3}(\lambda) P_{4}(\lambda),
\end{eqnarray}
where 
\begin{eqnarray}
    P_{1}(\lambda) &=& \lambda^{2}, \\
    P_{2}(\lambda) &=& \lambda (\lambda+2) (\lambda - 4 ), \\
    P_{3}(\lambda) &=& (\lambda^2-1)(\lambda^2-5), \\
    P_{4}(\lambda) &=& \lambda^{2} (\lambda+2).
\end{eqnarray}

The nontrivial integer factorization within the single quantum number sector indicates the existence of Hilbert space fragmentation. To further understand this connection, we derive the Krylov subspaces associated to each polynomial factor, namely $P_{1}(\lambda), P_{2}(\lambda), P_{3}(\lambda), P_{4}(\lambda)$ respectively. Obviously, the Krylov subspaces associated with $P_{1}(\lambda)$ are $\{ \vert \psi_{1} \rangle, \vert \psi_{2} \rangle \}$ because they are frozen states. $P_{2}(\lambda)$ is associated to the XX model integrable subspace. To illustrate this connection, we choose the unnormalized initial state $\vert \psi_{3} \rangle = \vert 100110011001 \rangle$, which yields
\begin{eqnarray}
    H^{0}\vert \psi_{3} \rangle &=& \vert \psi_{3} \rangle, \\
    H^{1} \vert \psi_{3} \rangle &=& -\vert \psi_{4} \rangle + \vert \psi_{5} \rangle, \\
    H^{2} \vert \psi_{3} \rangle &=& 6 \vert \psi_{3} \rangle + 2 (-\vert \psi_{4} \rangle +  \vert \psi_{5} \rangle) - 2 \vert \psi_{6} \rangle, \\
    H^{3} \vert \psi_{3} \rangle &=& 6 H^{1} \vert \psi_{3} \rangle + 2 H^{2} \vert \psi_{3} \rangle - 2 H^{1} \vert \psi_{6} \rangle = 8 H^{1} \vert \psi_{3} \rangle + 2 H^{2} \vert \psi_{3} \rangle,
\end{eqnarray}
which indicates a $n=3$ dimensional Krylov subspace with $v_{0} =0, v_{1} = 8, v_{2} = 2$ and thus corresponds to $P_{2}(\lambda)$. $P_{3}(\lambda)$ corresponds to the Krylov subspace generated by $\vert \psi_{7} \rangle = \vert 111100100100 \rangle$. We choose $\vert \psi_{7} \rangle$ as the initial state and then we have
\begin{eqnarray}
    H^{0} \vert \psi_{7} \rangle &=& \vert \psi_{7} \rangle, \\
    H^{1} \vert \psi_{7} \rangle &=& -\vert \psi_{8} \rangle - \vert \psi_{9} \rangle + \vert \psi_{10} \rangle, \\
    H^{2} \vert \psi_{7} \rangle &=& 3\vert \psi_{7} \rangle + 2\vert \psi_{11} \rangle, \\
    H^{3} \vert \psi_{7} \rangle &=& -5 \vert \psi_{8} \rangle - 3 \vert \psi_{9} \rangle + 5\vert \psi_{10} \rangle + 2 \vert \psi_{12} \rangle, \\
    H^{4} \vert \psi_{7} \rangle &=& 5H (-\vert \psi_{8}-\vert \psi_{9} \rangle + \vert \psi_{10} \rangle ) + 2 H \vert \psi_{9} \rangle + 2 H \vert    \psi_{12} \rangle \\ \nonumber
    &=& 5H^{2} \vert \psi_{7} \rangle - 2 \vert \psi_{7} \rangle + 2 \vert \psi_{11} \rangle \\ \nonumber
    &=&  5H^{2} \vert \psi_{7} \rangle  + H^{2} \vert \psi_{7} \rangle - 5 H^{0} \vert \psi_{7} \rangle \\ \nonumber 
    &=& 6 H^{2} \vert \psi_{7} \rangle - 5 H^{0} \vert \psi_{7} \rangle,
\end{eqnarray}
which indicates a $n=4$ Krylov subspace with $v_{0}=-5, v_{1} =0, v_{2} = 6, v_{3} = 0$ and thus corresponds to $P_{3}(\lambda)$. Since all the roots of $P_{4}(\lambda)$ are integers, they can be associated to integer eigenvectors namely $\vert \psi_{4} \rangle + \vert \psi_{5} \rangle$ (with eigenvalue $-2$), $\vert \psi_{8} \rangle + \vert \psi_{10} \rangle$ and $\vert \psi_{8} \rangle - \vert \psi_{9} \rangle + \vert \psi_{12} \rangle $ (both with eigenvalues 0). Therefore, we have demonstrated the equivalence between nontrivial integer factorization and the existence of Hilbert space fragmentation.

Similarly, we can also obtain the integer factorization in larger systems. For example, the analogue of $P_{3}$ with $L=16$ corresponds to a Krylov subspace of dimension 19 and the associated characteristic polynomial is shown in Eq.~(17) in the main text. If we directly work in a given parity sector of the p-h symmetry, 
\begin{eqnarray}
    \lambda^{10} - 25 \lambda^{8} + 188 \lambda^{6}  - 519 \lambda^{4} + 479 \lambda^{2} -72
\end{eqnarray}
corresponds to the Krylov subspace generated by $\vert 1111001001100100 \rangle + \vert 0000110110011011 \rangle$ and
\begin{eqnarray}
    \lambda (\lambda^8 - 15 \lambda^6 + 56 \lambda^4 -61 \lambda^2 + 15)
\end{eqnarray}
corresponds to the Krylov subspace generated by $\vert 1111001001100100 \rangle - \vert 0000110110011011 \rangle$.

\section{Nontrivial Krylov subspaces with momentum sector $k=\pi/2$}\label{app:pi2}
The connection between nontrivial integer characteristic polynomial factorization and the existence of Hilbert space fragmentation can be extended to general momentum sectors beyond just $k=0$ or $k=\pi$. Here, we showcase by focusing on $k= \pi/2$ momentum sector.

We consider the pair-hopping model and set $L=12$, $Q=4$, $Q_{e}=2$, $D=10$, $\kappa = 1$, which corresponds to the largest Hilbert subspace of dimension $9$ for this system size at $1/3$ filling. The momentum is $k = 2 \pi \kappa/ Q = \pi/2$. The Hilbert subspace is spanned by 
\begin{eqnarray}
    & \vert \psi_{1} \rangle = \vert 1 1 1 0 0 0 0 1 0 0 0 0 \rangle, \\ 
    & \vert \psi_{2} \rangle = \vert 1 1 0 0 0 0 0 0 0 0 1 1 \rangle, \\
    & \vert \psi_{3} \rangle = \vert 1 0 0 0 1 0 0 1 0 0 0 1 \rangle, \\
    & \vert \psi_{4} \rangle = \vert 1 1 0 1 0 0 1 0 0 0 0 0 \rangle, \\
    & \vert \psi_{5} \rangle = \vert 1 1 0 0 1 1 0 0 0 0 0 0 \rangle, \\
    & \vert \psi_{6} \rangle = \vert 1 0 1 1 0 1 0 0 0 0 0 0 \rangle, \\
    & \vert \psi_{7} \rangle = \vert 1 0 0 1 0 0 0 0 1 0 0 1 \rangle, \\
    & \vert \psi_{8} \rangle = \vert 1 0 0 0 0 1 0 1 0 0 1 0 \rangle, \\
    & \vert \psi_{9} \rangle = \vert 0 1 1 0 0 0 0 0 1 0 0 1 \rangle.
\end{eqnarray}
We note that the product state configuration should be understood as the superposition of all its symmetry partners in order to be invariant under the discrete translation symmetry. For example, the full expression of $\vert \psi_{4} \rangle$ is given by

\begin{eqnarray}
\vert \psi_{4} \rangle = &\vert 1 1 0 1 0 0 1 0 0 0 0 0 \rangle \equiv & \vert 1 1 0 1 0 0 1 0 0 0 0 0 \rangle + e^{i \frac{\pi}{2}} \vert 0 0 0 1 1 0 1 0 0 1 0 0 \rangle \nonumber\\
&& - e^{2 i \frac{\pi}{2}} \vert 1 0 0 0 0 0 1 1 0 1 0 0 \rangle + e^{3i \frac{\pi}{2}} \vert 1 0 0 1 0 0 0 0 0 1 1 0\rangle.
\end{eqnarray}
The minus sign in front of $\vert 1 0 0 0 0 0 1 1 0 1 0 0 \rangle$ is induced by the translation operator that translates all fermions by $L/Q=3$ sites. 
Within this subspace, the Hamiltonian matrix in the normalized basis is:
\begin{align}
    H = \begin{pmatrix}
     0 &  &  &  &  &  &  &  &  \\
      & 0 &  &  &  &  &  &  & \\
      &   & 0&  &  &  &  &  &\\
      &  &  & 0 & -1& 0 & -i & 0 & 0 \\
      &  & & -1  & 0 & -1 & 0 & 0 & -i \\
      &  & & 0  & -1& 0 & 0 & 0 & 0 \\
      & & &  i& 0 & 0 & 0 & i & -1 & \\
      & & & 0 &0 &0  & -i & 0 & 0 \\ 
      & & & 0 & i & 0 & -1 & 0 & 0
        \end{pmatrix},
\end{align} 
The characteristic polynomial within this single quantum number sector is 
\begin{eqnarray}
    P(\lambda) = P_{1}(\lambda) P_{2}(\lambda) P_{3}(\lambda) P_{4}(\lambda),
\end{eqnarray}
where 
\begin{eqnarray}
    & P_{1}(\lambda) = \lambda^{2}, \\ 
    & P_{2}(\lambda) = \lambda (\lambda^{2}-5), \\
    & P_{3}(\lambda) = \lambda^{2}-1, \\
    & P_{4}(\lambda) = \lambda^{2}.
\end{eqnarray}
The Krylov subspace associated with $P_{1}(\lambda)$ are $\{\vert \psi_{1} \rangle, \vert \psi_{2} \rangle \}$, where $\vert \psi_{1} \rangle$ and $\vert \psi_{2} \rangle $ are frozen states. To obtain the Krylov subspace associated with $P_{2}(\lambda)$, we choose the initial state $\vert \psi_{4} \rangle$, which yields
\begin{eqnarray}
     H^{0} \vert \psi_{4} \rangle &=& \vert \psi_{4} \rangle, \\
    H^{1} \vert \psi_{4} \rangle &=& -\vert \psi_{5} \rangle +i \vert \psi_{7} \rangle, \\
    H^{2} \vert \psi_{4} \rangle &=& 2 \vert \psi_{4} \rangle + \vert \psi_{6} \rangle + \vert \psi_{8} \rangle - 2i \vert \psi_{9} \rangle, \\
    H^{3} \vert \psi_{4} \rangle &=& 5 H^{1} \vert \psi_{4} \rangle,
\end{eqnarray}
which indicates a $n=3$ Krylov subspace with $v_{0} = 0$, $v_{1} = 5$, $v_{2}= 0$ and thus corresponds to $P_{2}(\lambda)$. To find the Krylov subspace associated with $P_{3}(\lambda)$, we choose $\vert \psi_{5} \rangle + i \vert \psi_{7} \rangle$ as the initial state. Note that this state is not a simple product state: Such states generating fragmentation are usually referred to as quantum Hilbert space fragmentation in the literature. Despite this difference, this initial state can be obtained using Eq.~(8) in the main text. We then have:
\begin{eqnarray}
    H^{0} (\vert \psi_{5} \rangle + i \vert \psi_{7} \rangle)  &=& (\vert \psi_{5} \rangle + i \vert \psi_{7} \rangle), \\
    H^{1} (\vert \psi_{5} \rangle + i \vert \psi_{7} \rangle)  &=&  -\vert \psi_{6} \rangle + \vert \psi_{8} \rangle, \\ 
    H^{2} (\vert \psi_{5} \rangle + i \vert \psi_{7} \rangle) &=& (\vert \psi_{5} \rangle + i \vert \psi_{7} \rangle),
\end{eqnarray}
which indicates a $n=2$ Krylov subspace with $v_{0}=1$ and thus corresponds to $P_{3}(\lambda)$. $P_{4}(\lambda)$ corresponds to the Krylov subspace generated by two eigenstates with zero energy, $\vert \psi_{3} \rangle$ and $\vert \psi_{6} \rangle + \vert \psi_{8} \rangle + i \vert \psi_{9} \rangle$.
\end{document}